\documentclass[%
superscriptaddress,
showpacs,preprintnumbers,
nofootinbib,
 amsmath,amssymb,
 aps,
 twocolumn,
prl,
nofootinbib
]{revtex4}
\usepackage{braket}
\usepackage{bm}
\usepackage{dcolumn}
\usepackage[dvipdfmx,final]{graphicx}
\input{colordvi.tex}
\usepackage{longtable}
\usepackage{hyperref}
\usepackage[utf8]{inputenc}

\def\bea{\begin{eqnarray}}
\def\eea{\end{eqnarray}}

\def\GeV{\,{\rm GeV}}
\def\MeV{\,{\rm MeV}}
\def\keV{\,{\rm keV}}
\def\eV{\,{\rm eV}}

\def\Mpc{\,{\rm Mpc}}

\def\vrel{v_{\rm rel}}
\def\semi{\sigma_{\rm semi}\vrel}
\def\invsemi{\sigma_{\rm inv} \vrel}
\def\eq{{\rm eq}}

\usepackage{color}
\usepackage{ulem}

\begin{document}
\preprint{CTPU-17-27}
\title{Self-heating dark matter via semi-annihilation}

\author{Ayuki Kamada}
\email{akamada@ibs.re.kr}
\affiliation{Center for Theoretical Physics of the Universe, Institute for Basic Science (IBS), Daejeon 34126, Korea}
\author{Hee Jung Kim}
\email{hyzer333@kaist.ac.kr}
\affiliation{Department of Physics, KAIST, Daejeon 34141, Korea}
\author{Hyungjin Kim}
\email{hjkim06@kaist.ac.kr}
\affiliation{Center for Theoretical Physics of the Universe, Institute for Basic Science (IBS), Daejeon 34126, Korea}
\affiliation{Department of Physics, KAIST, Daejeon 34141, Korea}
\affiliation{Department of Particle Physics and Astrophysics, Weizmann Institute of Science, Rehovot 7610001, Israel}
\author{Toyokazu Sekiguchi}
\email{sekiguti@ibs.re.kr}
\affiliation{Center for Theoretical Physics of the Universe, Institute for Basic Science (IBS), Daejeon 34126, Korea}
\affiliation{Research Center for the Early Universe (RESCEU), Graduate School of Science, The University of Tokyo, Tokyo 113-0033, Japan}
\date{\today}

\begin{abstract}
The freeze-out of dark matter (DM) depends on the evolution of the DM temperature. The DM temperature does not have to follow the standard model one, when the elastic scattering is not sufficient to maintain the kinetic equilibrium.
We study the temperature evolution of the semi-annihilating DM, where a pair of the DM particles annihilate into one DM particle and another particle coupled to the standard model sector.
We find that the kinetic equilibrium is maintained solely via semi-annihilation until the last stage of the freeze-out.
After the freeze-out, semi-annihilation converts the mass deficit to the kinetic energy of DM, which leads to non-trivial evolution of the DM temperature.
We argue that the DM temperature redshifts like radiation as long as the DM self-interaction is efficient.
We dub this novel temperature evolution as {\it self-heating}.
Notably, the structure formation is suppressed at subgalactic scales like keV-scale warm DM but with GeV-scale self-heating DM if the self-heating lasts roughly until the matter-radiation equality.
The long duration of the self-heating requires the large self-scattering cross section, which in turn flattens the DM density profile in inner halos.
Consequently, self-heating DM can be a unified solution to apparent failures of cold DM to reproduce the observed subgalactic scale structure of the Universe.
\end{abstract}

\pacs{95.35.+d} 

\maketitle

{\it Introduction} -- Particle dark matter (DM) is a fascinating possibility since it indicates new physics beyond the standard model (SM).
Such new physics should provide a mechanism that stabilizes a DM particle at least over the age of the Universe.
A simple approach is to introduce a $Z_{2}$ symmetry under which the SM particles and the DM particle have even and odd parity, respectively.
The $Z_{2}$ symmetry allows the DM particles to pair-annihilate into SM particles.
The observed DM mass density is reproduced when the pair-annihilation cross section is around the electroweak scale.
Known as the weakly interacting massive particle (WIMP), such an electroweak-scale DM candidate can be embedded in well-motivated new physics such as supersymmetry.
However, despite the extensive efforts so far, no signal of WIMP has been found\,\cite{Arcadi:2017kky, Roszkowski:2017nbc}.
In light of this situation, it appears plausible to revisit the longevity of a DM particle and associated symmetries beyond $Z_{2}$.

Other symmetries, such as $Z_{3}$ and SU$(N)$, can also guarantee the stability of a DM particle.
In the $Z_3$ case, the DM relic abundance can be determined by {\it semi-annihilation}\,\cite{Hambye:2008bq, Hambye:2009fg, Arina:2009uq, DEramo:2010keq},
\bea
\chi \chi \to \chi \phi \,, 
\label{eq:semi}
\eea
where $\chi$ represents the DM particles and $\phi$ is a particle coupled to the SM sector.
One may consider dark meson states in a hidden confining sector as DM. 
Their stability is protected by a flavor SU$(N)$ symmetry.
This is called the strongly interacting massive particle (SIMP).
Their relic abundance is determined by the $3 \to 2$ process\,\cite{Carlson:1992fn, Hochberg:2014dra, Hochberg:2014kqa}.
In these cases, the elastic scattering of a DM particle with a SM one is not directly related with semi-annihilation or the $3 \to 2$ process.%
\footnote{In the WIMP case, the elastic scattering always accompanies the pair-annihilation due to the crossing symmetry.}
When the elastic scattering is inefficient, the DM temperature is not equilibrated with the SM one; i.e., the kinetic equilibrium is not maintained.
On the other hand, the relic abundance heavily depends on the evolution of the DM temperature.
In the SIMP scenario, if the DM sector is isolated from the SM sector, the entropy conservation infers that the DM number density redshifts only adiabatically\,\cite{Carlson:1992fn}.
The strength of the elastic scattering can dominantly determine the relic abundance\,\cite{Kuflik:2015isi, Kuflik:2017iqs}.

In this {\it Letter}, for the first time, we investigate how the semi-annihilating DM number density evolves when the elastic scattering is inefficient.
We argue that as long as the DM self-interaction is efficient, the temperature redshifts inversely proportional to the scale factor, $T_{\chi} \propto 1/a$, even after the freeze-out.
We dub this novel temperature evolution as {\it self-heating}.

Self-heating DM reconciles apparent failures of cold dark matter (CDM), which is represented by the aforementioned thermal WIMP, to reproduce the subgalactic scale structure of the Universe\,\cite{Bullock:2017xww}.%
\footnote{Baryonic feedback may play an important role in solving the problems\,\cite{Dutton:2015nvy, Sawala:2015cdf, Wetzel:2016wro}.}
For example, CDM appears to overpredict the number of subgalactic halos, which is known as the {\it missing satellite problem}\,\cite{Moore:1999nt, Kravtsov:2009gi}.
Warm dark matter (WDM) with keV-scale DM mass has been intensively investigated as a possible solution\,\cite{Bode:2000gq, Lovell:2011rd}.
In the self-heating scenario, the relatively high temperature suppresses the formation of the subgalactic-scale structure like keV-scale WDM but with GeV-scale DM mass.
Since a large self-scattering cross section is essential for a sufficiently long duration of the self-heating regime, self-heating DM is also a natural solution to the {\it core-cusp problem}.
While the CDM predicts the steep density profile in inner halos\,\cite{Moore:1999gc, deBlok:2009sp}, 
the self-scattering cross section as large as $\sigma_{\rm self} / m_{\chi} \sim 1 \, {\rm cm^{2} / g}$ flattens the inner density profile to be compatible with the observation\,\cite{Spergel:1999mh, Tulin:2017ara}.%
\footnote{The constraints from galaxy clusters may favor a velocity-dependent self-scattering cross section\,\cite{Peter:2012jh, Markevitch:2003at, Randall:2007ph}.}
\\

\begin{figure}
\centering
\includegraphics[scale=0.2]{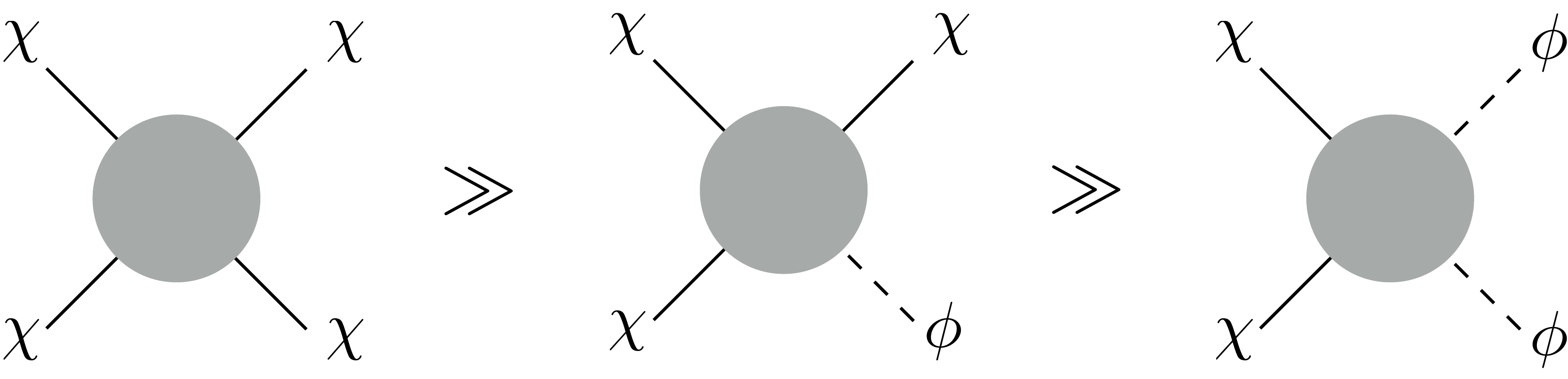}
\caption{A schematic figure showing the DM interactions under consideration and their relative strength.
We assume that the self-interaction is the strongest interaction, while the pair-annihilation is the weakest and irrelevant.}
\label{fig:int_hier}
\end{figure}

{\it Freeze-out} -- We consider that a DM particle ($\chi$) with mass $m_{\chi}$ interacts with a light particle ($\phi$) through semi-annihilation as in Eq.\,\eqref{eq:semi}.
We assume that $\phi$ is in thermal equilibrium with the SM plasma, $T_{\phi} = T$, and also that the DM self-interaction is efficient during and after the freeze-out.
On the other hand, we ignore other interactions including the pair-annihilation and the elastic scattering of DM with a SM particle.
Figure\,\ref{fig:int_hier} schematically describes our setup.
One of theoretical setups realizing this hierarchy among the cross sections is a SIMP model assisted by an axion-like particle (ALP)\,\cite{Kamada:2017tsq}.
In this model, the ALP is identified as $\phi$, and the interaction strength of processes involving a larger number of $\phi$'s is suppressed by a larger power of the ALP decay constant.

In the above setup, the DM phase space distribution with the energy of $E_{\chi}$ follows the Maxwell-Boltzmann one:
\bea
f_{\chi} = \frac{n_{\chi}}{n_{\chi}^{\eq} (T_{\chi})} \exp\left(- E_{\chi} / T_{\chi} \right) \,,
\label{eq:df}
\eea
where $n_{\chi}^{\eq} = m_{\chi}^2 T_{\chi} / (2 \pi^{2}) \, K_{2} (m_{\chi} / T_{\chi})$ with $K_{2}$ being the second-order modified Bessel function of the second kind.
For convenience, we introduce the DM yield, $Y_{\chi} = n_{\chi} / s$, in addition to $x_{\chi} = m_{\chi} / T_{\chi}$, where $s = (2 \pi^{2} / 45) \, g_{*} T^{3}$ and $g_{*}$ is the effective number of relativistic degrees of freedom.
From the Boltzmann equation of $f_{\chi}$, we derive the evolution equations of $Y_{\chi}$ and $x_{\chi}$ as%
\footnote{For Eqs.\,\eqref{eq:yield_2} and \eqref{eq:T_2}, we integrate the Boltzmann equation for $f_{\chi}$ over the phase space with and without the weight of $E_\chi$.  
Note that the DM self-interaction does not contribute to these evolution equations since it conserves the number and the energy of DM.
The detailed derivation will be given in Ref.\,\cite{KKKSip}.}

\bea
\frac{d}{dx}Y_{\chi} &=&
- \frac{\lambda}{x^{2}}Y_{\chi} \left[ Y_{\chi} - Y_{\chi}^{\eq}(x_{\chi}) {\cal J}(x_{\chi}, x) \right] \,,
\label{eq:yield_2}
\\
 \sigma^{2}_{E_{\chi}/m_{\chi}} \frac{d}{dx}x_{\chi} &=&
\frac{3}{x} \frac{1}{x_{\chi}} + 
\frac{\bar{\lambda} }{x^{2}} \left[ Y_{\chi} - Y_{\chi}^\eq(x_{\chi}) {\cal K}(x_{\chi},x) \right] \,,
\label{eq:T_2}
\eea
where $x = m_{\chi} / T$ and $\sigma^{2}_{E_{\chi}/m_{\chi}} = \langle E_{\chi}^{2} / m_{\chi}^{2} \rangle_{T_{\chi}}- \langle E_{\chi} / m_{\chi} \rangle_{T_{\chi}}^{2}$.
The angle bracket subscripted with a temperature denotes the thermal average of $\chi$.
Hereafter, we set $g_{*}$ to be constant, specifically $g_{*} =60$ for numerical results.
The dimensionless reaction rates, $\lambda$ and $\bar{\lambda}$, are given by
\bea
\lambda &=&
\frac{x s\langle \semi \rangle_{T_{\chi}, T_{\chi}}}{2H} \,,
\label{eq:lambda}
\\
\bar{\lambda} &=&
\frac{x s \langle \Delta E / m_{\chi} \, \invsemi \rangle_{T_{\chi}, T_{\phi} = T_{\chi}} } { H } 
\frac{n_{\phi}^{\eq}(T_{\phi} = T_{\chi})}{n_{\chi}^{\eq}(T_{\chi})} \,,
\label{eq:lambdab}
\eea
where $\Delta E = E_{\phi} - \langle E_{\chi}\rangle_{T_{\chi}}$ and $H^{2} = g_{*} \pi^{2} T^{4} / (90m_{\rm pl}^{2})$ with $m_{\rm pl}$ being the reduced Planck mass.%
\footnote{Note that $T_{\phi} = T_{\chi}$ in Eq.~\ref{eq:lambdab} appears as a result of the {\it detailed balance}.}
The cross sections of the forward and backward semi-annihilations are denoted by $\semi$ and $\invsemi$, respectively.
The angle bracket subscripted with two temperatures denotes the thermal average over the two initial particles: $\chi \chi$ for $\semi$ and $\chi \phi$ for $\invsemi$.
In addition, we define
\bea
{\cal J}(x_{\chi}, x) &\equiv& 
\frac{n_{\phi}^\eq(T_{\phi} = T)}{n_{\phi}^\eq(T_{\phi} = T_{\chi})}
\frac{\langle \invsemi \rangle_{T_{\chi}, T_{\phi} = T}}{\langle \invsemi \rangle_{T_{\chi}, T_{\phi} = T_{\chi}}} \,,
\\
{\cal K}(x_{\chi}, x) &\equiv&
\frac{n_{\phi}^\eq(T_{\phi} = T)}{n_{\phi}^\eq (T_{\phi} = T_{\chi})}
\frac{\langle \Delta E \, \invsemi \rangle_{T_{\chi}, T_{\phi} = T}}{\langle \Delta E \, \invsemi\rangle_{T_{\chi}, T_{\phi} = T_{\chi}}} \,.
\eea
Note that ${\cal J} = {\cal K} = 1$ when $T_{\chi} = T$.

\begin{figure}[t]
\centering
\includegraphics[scale=0.45]{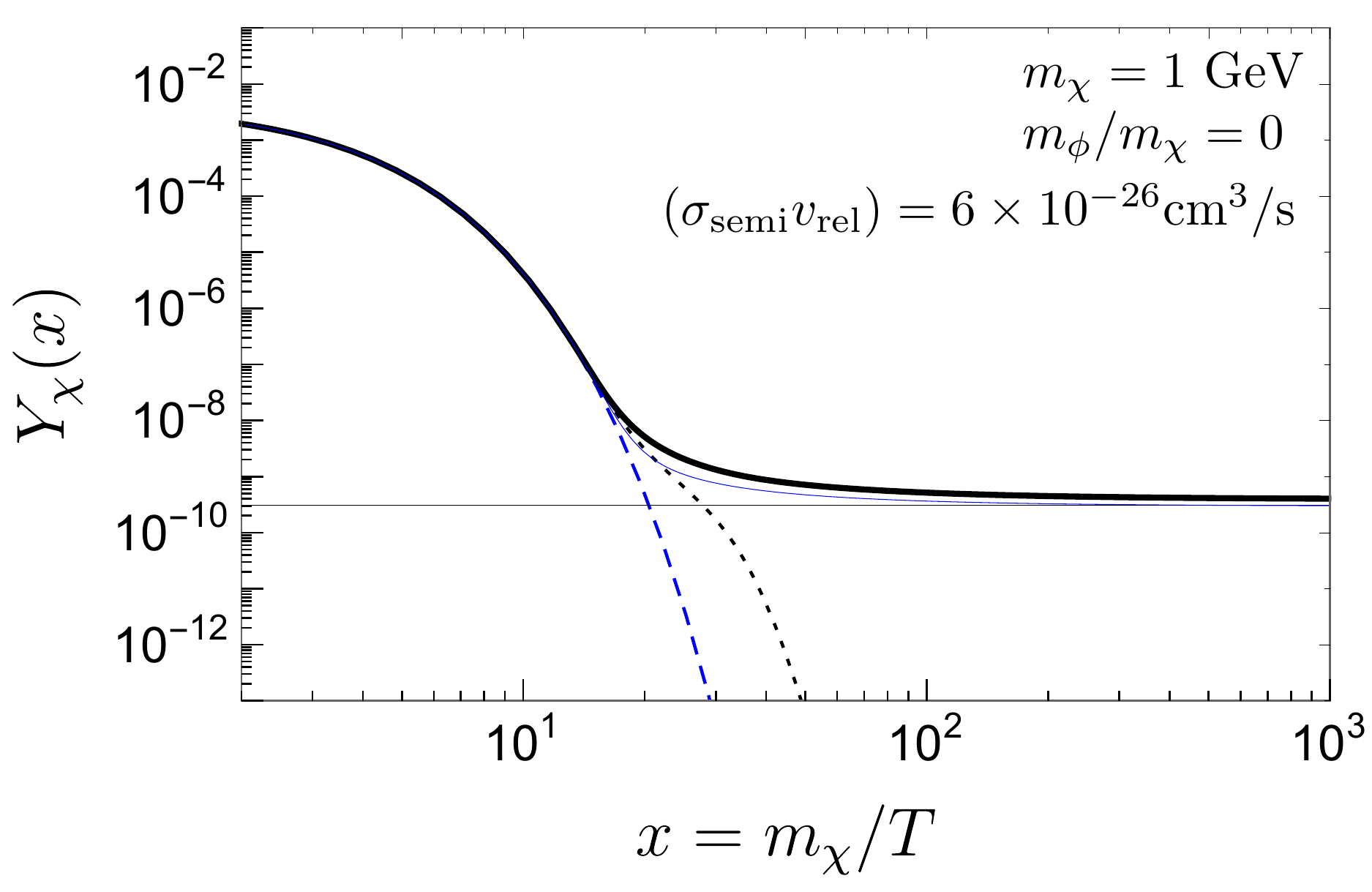}
\caption{Evolution of the DM yield.
The black solid line is $Y_{\chi}$, while the black dotted line is $Y_{\chi}^\eq(x_{\chi}) {\cal J}(x_{\chi},x)$.
The blue lines are the same except that $T_{\chi} = T_{\phi}$ is kept by hand, whose asymptotic value is indicated by the horizontal gray line.
}
\label{fig:yield}
\end{figure}

We present the evolution of the DM yield in Fig.\,\ref{fig:yield}, by following the co-evolution of $Y_{\chi}$ and $x_{\chi}$ numerically.
Hereafter, numerical results assume that the semi-annihilation invariant amplitude is constant, i.e., the $s$-wave semi-annihilation in the non-relativistic limit.
We also show the evolution of the DM yield when $T_{\chi} = T$, i.e., when the elastic scattering of a DM particle with a SM particle is efficient.
In both the cases, the DM yield follows its equilibrium value, $Y_{\chi} \simeq Y_{\chi}^\eq(x_{\chi}) {\cal J}(x_{\chi}, x)$, until $x \sim 20$.
Then, semi-annihilation freezes out, and the comoving number density is virtually conserved. 
There is a $\sim 30\%$ difference in the DM yield between the two cases.
The freeze-out process is delayed in the co-evolution case since the equilibrium value (black dotted line) is enhanced by the increasing DM temperature.

\begin{figure}[t]
\centering
\includegraphics[scale=0.4]{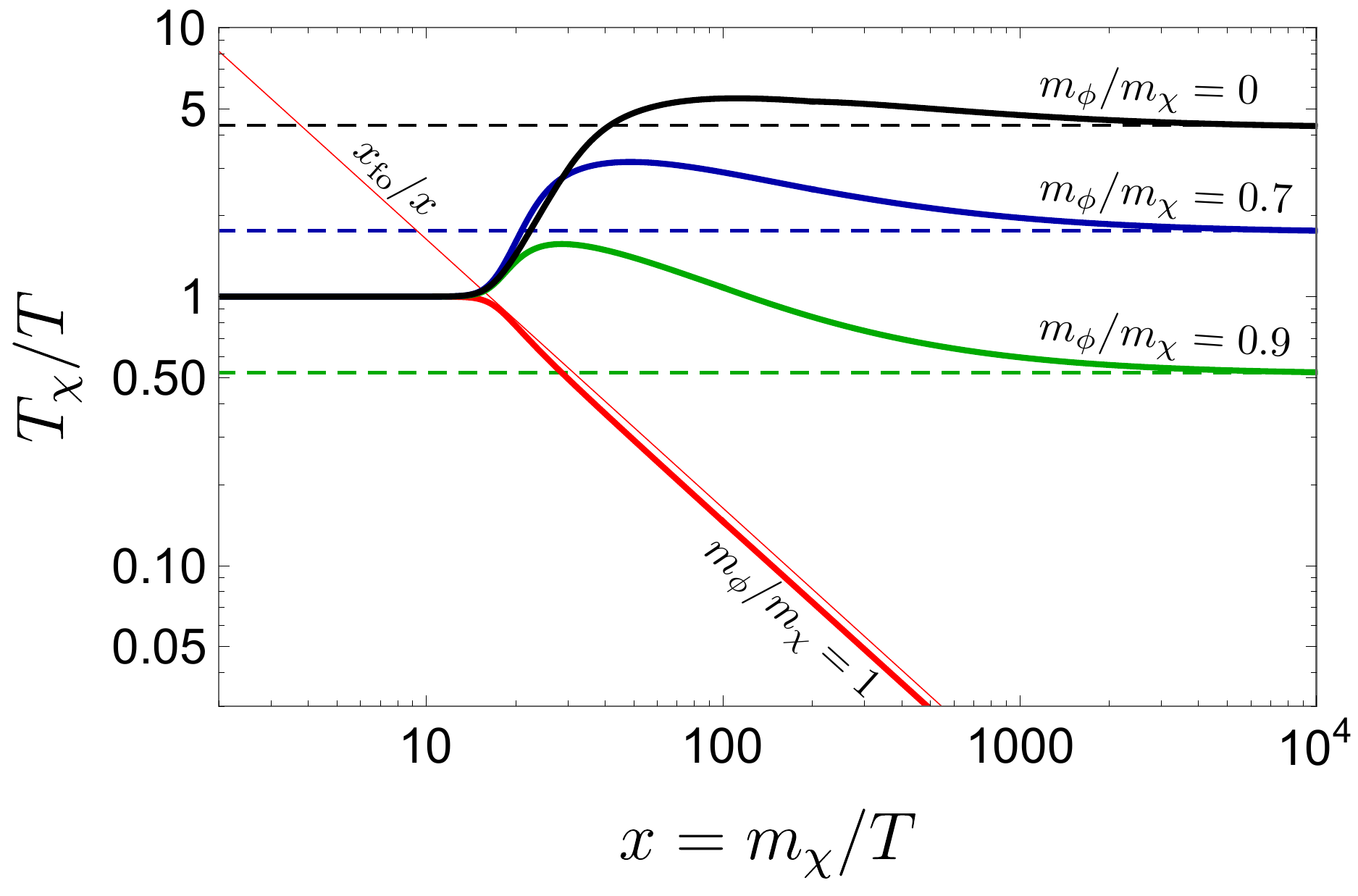}
\caption{Evolution of the DM temperature normalized by the SM one.
Different colors represent different values of $m_{\phi} / m_{\chi}$, while the invariant amplitude is fixed to reproduce $(\semi) = 6 \times 10^{-26} \, {\rm cm^{3} / s}$ for $m_{\phi} / m_{\chi} = 0$ (Fig.\,\ref{fig:yield}).
The solid lines are the numerical results, while the dashed lines are the analytic estimations of the asymptotic value, Eq.\,\eqref{eq:asym_T_ratio}.
The thin red line shows the adiabatic cooling: $T_{\chi} / T = x_{\rm fo} / x$ (for the definition of $x_{\rm fo}$, see footnote 6).}
\label{fig:T_ratio}
\end{figure}

In contrast to the DM number density, the DM temperature exhibits a unique behavior during and after the freeze-out.
Figure\,\ref{fig:T_ratio} shows the evolution of $T_{\chi} / T \, (= x / x_{\chi})$.
The DM particles remain in kinetic equilibrium with $\phi$'s and thus with SM particles until the freeze-out.
Semi-annihilation on its own suffices to equilibrate the DM and SM sectors kinetically as well as chemically.
After the freeze-out, the DM temperature begins to deviate from the SM one.
One may expect that the DM temperature adiabatically evolves as $T_{\chi} \propto 1 / a^{2}$ due to the expansion of the Universe.
However, in fact, the temperature ratio increases for a duration of $\Delta x \sim {\cal O}(10)$.
The increase in the DM temperature prompts the backward semi-annihilation process and slows the progress of the freeze-out, as we see in Fig.\,\ref{fig:yield}.
Eventually, the DM temperature scales as $T_{\chi} \propto 1 / a$ like radiation.

The reason behind the unique evolution of the DM temperature is that  semi-annihilation kinematically boosts the non-relativistic DM particles.
The Lorentz factor that the DM particle achieves through semi-annihilation is
\bea
\gamma = \frac{E_{\chi}}{m_{\chi}} = \frac{5}{4} \left( 1 - \frac{m_{\phi}^{2}}{5m_{\chi}^{2}} \right) \,.
\eea
As long as the DM particles self-interact with each other efficiently, the kinetic energy of the boosted DM particle is quickly redistributed over all the DM particles, which virtually heats DM.
Thus the self-heating regime, $T_{\chi} \propto 1 / a$, lasts until the self-interaction freezes out.

One can derive the behavior of the DM temperature after the freeze-out as follows.
After the freeze-out, Eqs.\,\eqref{eq:yield_2} and \eqref{eq:T_2} are approximated as
\bea
\frac{d}{dx}Y_{\chi} &\approx& 
- \frac{\lambda}{x^{2}} Y^{2}_{\chi} \,,
\label{eq:yield_non_rel}
\\
x \frac{d}{dx}\left(\frac{x_{\chi}}{x} \right) &\approx& 
\frac{x_{\chi}}{x}
+
\frac{2}{3} \bar{\lambda} \, Y_{\chi} \left( \frac{x_{\chi}}{x} \right)^{2} \,,
\label{eq:T_non_rel}
\eea 
where we neglect all the terms originating from the backward semi-annihilation process, and use the non-relativistic expression of the variance of the DM energy, $\sigma^{2}_{E/m_{\chi}} \approx 3 / (2 x_{\chi}^{2})$.
Assuming the $s$-wave semi-annihilation, i.e., constant $\lambda$, one can analytically solve the evolution equation of the DM yield and find its asymptotic value as
\bea
Y_{\chi,\infty} \approx x_{\rm fo} / \lambda (x_{\rm fo}) \,,
\eea
where $x_{\rm fo} = m_\chi / T_{\rm fo}$ with $T_{\rm fo}$ being the freeze-out temperature.%
\footnote{In this Letter, we determine $x_{\rm fo}$ by equating $x_{\rm fo} / \lambda$ [$2 x_{\rm fo} / \lambda (x_{\rm fo})$ for $m_{\phi} / m_{\chi} = 1$ since $(\semi) \propto v_{\rm rel}$] to the numerically obtained value of $Y_{\chi,\infty}$.
Note that a simple definition of $x_{\rm fo}$ found in the literature\,\cite{DEramo:2010keq} may result in a marginal difference\,\cite{KKKSip}.}
In addition, as one can see from Eqs.\,\eqref{eq:lambda} and \eqref{eq:lambdab}, $\bar{\lambda}$ is approximated as 
\bea
\bar{\lambda} \approx
- (\gamma -1) \lambda 
\approx
- (\gamma -1)  \frac{x_{\rm fo}}{Y_{\chi,\, \infty}} \,,
\eea
in the non-relativistic limit.
Using these expressions, one finds the evolution equation of the DM temperature as
\bea
x \frac{d}{d x} \left( \frac{x_{\chi}}{x}\right) &\approx& 
\frac{x_{\chi}}{x} -
(\gamma-1)
\frac{2 x_{\rm fo}}{3} \left( \frac{Y_{\chi}}{Y_{\chi,\, \infty}} \right) \left( \frac{x_{\chi}}{x} \right)^{2} \,.
\eea
Here, the first term on the right hand side represents the adiabatic cooling due to the expansion of the Universe, while the second term represents the heating through semi-annihilation.

\begin{figure}[!t]
\centering
\includegraphics[scale=0.37]{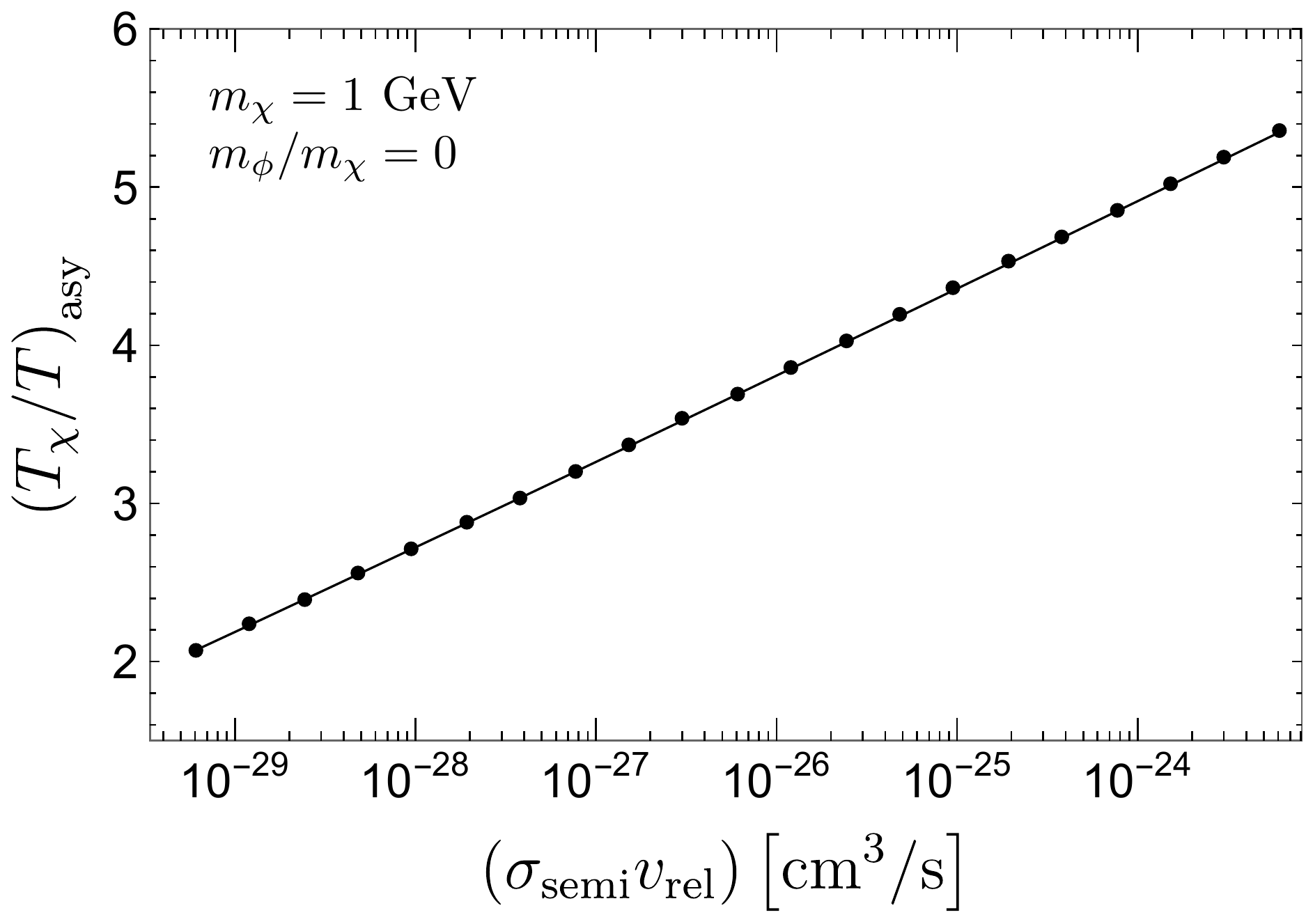}
\caption{Asymptotic value of the temperature ratio $\left( T_{\chi} / T \right)_{\rm asy}$ as a function of the semi-annihilation cross section in the non-relativistic limit.
The dots are the numerical results ($T_{\chi} / T$ at $x = 10^{5}$), while the solid line is the analytic estimation given in Eq.\,\eqref{eq:asym_T_ratio}.
Since $\left( T_{\chi} / T \right)_{\rm asy}$ is inversely proportional to the freeze-out temperature, it increases logarithmically as the cross section increases.}
\label{fig:asymp_T_ratio}
\end{figure}

Just after the freeze-out, the heating through semi-annihilation dominates the adiabatic cooling, and thus $T_{\chi} / T$ increases.
As the DM yield approaches its asymptotic value $Y_{\chi,\, \infty}$, the cooling and the heating balance with each other and $T_{\chi} / T$ becomes constant.
We find the asymptotic value of $T_{\chi} / T$ as
\bea
\left( \frac{T_{\chi}}{T} \right)_{\rm asy} \approx (\gamma - 1) \frac{2 x_{\rm fo}}{3} \,,
\label{eq:asym_T_ratio}
\eea
which is depicted with the dashed line in Fig.\,\ref{fig:T_ratio}.
This expression indicates that the asymptotic temperature ratio is inversely proportional to the freeze-out temperature.
Since the freeze-out temperature logarithmically depends on the semi-annihilation cross section, the temperature ratio should also do so.
In Fig.~\ref{fig:asymp_T_ratio}, we show the numerically obtained value of $T_{\chi} / T$ at $x = 10^{5}$ as a function of the cross section in the non-relativistic limit.
As expected, the analytic estimation given in Eq.~\eqref{eq:asym_T_ratio} agrees with the numerical result.
\\

{\it Impacts on structure formation} -- The self-heating regime lasts until the self-interaction freezes out when the temperature of the Universe becomes 
\bea
\label{eq:tself}
T_{\rm self} \simeq
1 \, {\rm eV} 
\left(\frac{1 \, {\rm cm}^{2} / {\rm g}}{\sigma_{\rm self} / m_{\chi}} \right)^{2/3} 
\left(\frac{m_{\chi} }{ 1\GeV} \right)^{1/3} \left( \frac{T_{\chi}}{T} \right)_{\rm asy}^{-1/3} \,.
\eea
After the freeze-out of the self-interaction, the energy of the boosted DM particles cannot be efficiently transferred to the others.
The boosted DM particles start to freely stream at the speed of light; i.e., they behave as a hot component of DM.
On the other hand, the fraction of such boosted DM particles is as tiny as
\bea
\int^{t_{\rm now}}_{t_{\rm self}} dt \, \langle \sigma_{\rm semi} \vrel \rangle n_{\chi} \simeq 
2 \times 10^{-8} 
\left(  \frac{T_{\rm self} }{ 1 \eV} \right) 
\left( \frac{50 \MeV}{ T_{\rm fo}} \right) \,,
\eea
where $t_{\rm now}$ and $t_{\rm self}$ denote the cosmic time at present and at the freeze-out of the self-interaction, respectively.
The impacts of the boosted DM particles on the structure formation are thus negligible.
Meanwhile, the majority of the DM particles follow the Maxwell-Boltzmann distribution with the DM temperature being adiabatically cooled, $T_{\chi} \propto 1/a^{2}$, and behave as a warm component of DM.

Such warmness of DM leaves a cutoff in the resultant matter power spectrum.
One way to estimate the cutoff scale is to compute the Jeans scale at the matter-radiation equality\,\cite{Kamada:2013sh}.%
\footnote{It is defined as $k_{\rm J} = a \sqrt{ 4 \pi G \rho_{\rm m} / \langle \vec{v}^{2} \rangle } {\big |}_{a_{\eq}}$, where $G$ is the Newton constant, $\rho_{\rm m}$ is the mass density of the whole matter, and $\langle \vec{v}^2 \rangle$ is the variance of the DM velocity.}
In the self-heating DM scenario, we find
\bea
\label{eq:kJ}
k_{\rm J}
\simeq 
220 \Mpc^{-1} 
\max\left( 1 , \sqrt{\frac{T_{\rm self} }{ T_{\eq} } } \right)
\left({ \frac{ m_{\chi} }{ 1\GeV}  }\right)^{1/2}
\left( \frac{T_{\chi}}{T} \right)_{\rm asy}^{-1/2} \,.
\eea
In the thermal WDM model,%
\footnote{There, DM particles follow the Fermi-Dirac distribution with two spin degrees of freedom.
The temperature is determined such that its mass density reproduces the observed value for a given mass.}
$k_{\rm J} \simeq 210 \Mpc^{-1} \left(m_{\rm wdm} / 6 \, {\rm keV} \right)^{4/3}$ with $m_{\rm wdm}$ being the thermal WDM mass\,\cite{Kamada:2013sh}.
It follows that the semi-annihilating GeV-scale DM suppresses the subgalactic-scale structure formation like keV-scale WDM.

The suppression of the subgalactic-scale structure formation is constrained by measurements of the Lyman-$\alpha$ forest.
The thermal WDM mass is constrained as $m_{\rm wdm} \geq 5.3\keV$ at the 2$\sigma$ level\,\cite{Irsic:2017ixq, Yeche:2017upn}.
By comparing the above cutoff scales estimated by $k_{\rm J}$, we could translate the constraint on $m_{\rm wdm}$ to $m_{\chi}$ and $T_{\rm self}$ as
\bea
\frac{m_{\rm wdm}}{ 5.3\keV} \simeq 
\alpha
\left(\frac{m_\chi}{1 \GeV} \right)^{3/8} 
\max\left( 1 , \sqrt{\frac{T_{\rm self} }{ T_{\eq} } } \right)^{3/4}
\left(\frac{T_\chi}{T} \right)_{\rm asy}^{-3/8} \,.
\eea 
Here we introduce $\alpha \simeq 3$ so that the two models exhibit the same cutoff in the numerically calculated linear matter power spectrum~\cite{KKKSip}.

From Eqs.~\eqref{eq:tself} and \eqref{eq:kJ}, one can see that the suppression of the subgalactic-scale structure formation is interrelated with the large self-scattering cross section flattening the DM density profile in inner halos.
Thus self-heating DM solves the missing satellite problem and the core-cusp problem in one stroke.
This also provides an intriguing future prospect.
A larger number of kinematically resolved dwarf galaxies will pin down the self-interaction strength\,\cite{Valli:2017ktb}.
One can examine whether it is interrelated with the self-heating regime of DM by probing the smaller-scale structure of the Universe through the 21\,cm-line fluctuations due to minihalos\,\cite{Sekiguchi:2014wfa} and the flux ratios in strong gravitational lenses\,\cite{Gilman:2017voy}.
\\

{\it Implications for particle physics model building} -- We studied an unexplored region of the semi-annihilating DM, where the DM interactions are hierarchical as depicted in Fig.~\ref{fig:int_hier}.
They arise from different interaction terms in a Lagrangian and thus are independent in general, e.g., see a simple $Z_{3}$ model~\cite{DEramo:2010keq} and an extension of the SIMP model with an ALP~\cite{Kamada:2017tsq}.
We expect that the self-heating regime takes place even if semi-annihilation exists but does not dominate the DM freeze-out.
The heating through semi-annihilation compensates the adiabatic cooling by the expansion of the Universe and leads to constant $T_{\chi} / T$ after the usual kinetic decoupling.

While we leave a concrete particle physics realization of self-heating DM for future work~\cite{KKKSip}, the cosmological and astrophysical implications of $\phi$ are worth mentioning.
The energetic $\phi$'s are continuously produced by semi-annihilation even after the freeze-out.
Subsequent decay of produced $\phi$ into SM particles is subject to indirect detection searches.
The constraint on semi-annihilating DM mass is similar to that of the WIMP: $m_{\chi} > 10 \text{--} 100 \, {\rm GeV}$, depending on the decay products (see Refs.\,\cite{Arcadi:2017kky, Roszkowski:2017nbc} and references therein).
Consequently, in the setup of Ref.\,\cite{Kamada:2017tsq}, the large self-scattering cross section cannot be achieved within the bound, unless $m_\phi$ and $m_\chi$ are degenerate.
On the other hand, if $\phi$ is accommodated in a hidden sector and is thermalized with {\it dark radiation}, the indirect detection constraint is evaded.
In such a case, $T_{\phi}$ follows the hidden sector temperature, $T_{\rm d}$, and one needs to take account of $T_{\rm d} \neq T$ and take care of the dark radiation contribution to the effective number of neutrino species, which is constrained as $\Delta N_{\rm eff}<0.6 \; (2\sigma)$\,\cite{Ade:2015xua}.
\\

{\it Acknowledgements} -- This work was supported by Institute for Basic Science under the project code, IBS-R018-D1.

\bibliography{self-heating}

\begin{thebibliography}{35}
\expandafter\ifx\csname natexlab\endcsname\relax\def\natexlab#1{#1}\fi
\expandafter\ifx\csname bibnamefont\endcsname\relax
  \def\bibnamefont#1{#1}\fi
\expandafter\ifx\csname bibfnamefont\endcsname\relax
  \def\bibfnamefont#1{#1}\fi
\expandafter\ifx\csname citenamefont\endcsname\relax
  \def\citenamefont#1{#1}\fi
\expandafter\ifx\csname url\endcsname\relax
  \def\url#1{\texttt{#1}}\fi
\expandafter\ifx\csname urlprefix\endcsname\relax\def\urlprefix{URL }\fi
\providecommand{\bibinfo}[2]{#2}
\providecommand{\eprint}[2][]{\url{#2}}

\bibitem[{\citenamefont{Arcadi et~al.}(2018)\citenamefont{Arcadi, Dutra, Ghosh,
  Lindner, Mambrini, Pierre, Profumo, and Queiroz}}]{Arcadi:2017kky}
\bibinfo{author}{\bibfnamefont{G.}~\bibnamefont{Arcadi}},
  \bibinfo{author}{\bibfnamefont{M.}~\bibnamefont{Dutra}},
  \bibinfo{author}{\bibfnamefont{P.}~\bibnamefont{Ghosh}},
  \bibinfo{author}{\bibfnamefont{M.}~\bibnamefont{Lindner}},
  \bibinfo{author}{\bibfnamefont{Y.}~\bibnamefont{Mambrini}},
  \bibinfo{author}{\bibfnamefont{M.}~\bibnamefont{Pierre}},
  \bibinfo{author}{\bibfnamefont{S.}~\bibnamefont{Profumo}}, \bibnamefont{and}
  \bibinfo{author}{\bibfnamefont{F.~S.} \bibnamefont{Queiroz}},
  \bibinfo{journal}{Eur. Phys. J.} \textbf{\bibinfo{volume}{C78}},
  \bibinfo{pages}{203} (\bibinfo{year}{2018}), \eprint{1703.07364}.

\bibitem[{\citenamefont{Roszkowski et~al.}(2017)\citenamefont{Roszkowski,
  Sessolo, and Trojanowski}}]{Roszkowski:2017nbc}
\bibinfo{author}{\bibfnamefont{L.}~\bibnamefont{Roszkowski}},
  \bibinfo{author}{\bibfnamefont{E.~M.} \bibnamefont{Sessolo}},
  \bibnamefont{and}
  \bibinfo{author}{\bibfnamefont{S.}~\bibnamefont{Trojanowski}}
  (\bibinfo{year}{2017}), \eprint{1707.06277}.

\bibitem[{\citenamefont{Hambye}(2009)}]{Hambye:2008bq}
\bibinfo{author}{\bibfnamefont{T.}~\bibnamefont{Hambye}},
  \bibinfo{journal}{JHEP} \textbf{\bibinfo{volume}{01}}, \bibinfo{pages}{028}
  (\bibinfo{year}{2009}), \eprint{0811.0172}.

\bibitem[{\citenamefont{Hambye and Tytgat}(2010)}]{Hambye:2009fg}
\bibinfo{author}{\bibfnamefont{T.}~\bibnamefont{Hambye}} \bibnamefont{and}
  \bibinfo{author}{\bibfnamefont{M.~H.~G.} \bibnamefont{Tytgat}},
  \bibinfo{journal}{Phys. Lett.} \textbf{\bibinfo{volume}{B683}},
  \bibinfo{pages}{39} (\bibinfo{year}{2010}), \eprint{0907.1007}.

\bibitem[{\citenamefont{Arina et~al.}(2010)\citenamefont{Arina, Hambye, Ibarra,
  and Weniger}}]{Arina:2009uq}
\bibinfo{author}{\bibfnamefont{C.}~\bibnamefont{Arina}},
  \bibinfo{author}{\bibfnamefont{T.}~\bibnamefont{Hambye}},
  \bibinfo{author}{\bibfnamefont{A.}~\bibnamefont{Ibarra}}, \bibnamefont{and}
  \bibinfo{author}{\bibfnamefont{C.}~\bibnamefont{Weniger}},
  \bibinfo{journal}{JCAP} \textbf{\bibinfo{volume}{1003}}, \bibinfo{pages}{024}
  (\bibinfo{year}{2010}), \eprint{0912.4496}.

\bibitem[{\citenamefont{D'Eramo and Thaler}(2010)}]{DEramo:2010keq}
\bibinfo{author}{\bibfnamefont{F.}~\bibnamefont{D'Eramo}} \bibnamefont{and}
  \bibinfo{author}{\bibfnamefont{J.}~\bibnamefont{Thaler}},
  \bibinfo{journal}{JHEP} \textbf{\bibinfo{volume}{06}}, \bibinfo{pages}{109}
  (\bibinfo{year}{2010}), \eprint{1003.5912}.

\bibitem[{\citenamefont{Carlson et~al.}(1992)\citenamefont{Carlson, Machacek,
  and Hall}}]{Carlson:1992fn}
\bibinfo{author}{\bibfnamefont{E.~D.} \bibnamefont{Carlson}},
  \bibinfo{author}{\bibfnamefont{M.~E.} \bibnamefont{Machacek}},
  \bibnamefont{and} \bibinfo{author}{\bibfnamefont{L.~J.} \bibnamefont{Hall}},
  \bibinfo{journal}{Astrophys. J.} \textbf{\bibinfo{volume}{398}},
  \bibinfo{pages}{43} (\bibinfo{year}{1992}).

\bibitem[{\citenamefont{Hochberg et~al.}(2014)\citenamefont{Hochberg, Kuflik,
  Volansky, and Wacker}}]{Hochberg:2014dra}
\bibinfo{author}{\bibfnamefont{Y.}~\bibnamefont{Hochberg}},
  \bibinfo{author}{\bibfnamefont{E.}~\bibnamefont{Kuflik}},
  \bibinfo{author}{\bibfnamefont{T.}~\bibnamefont{Volansky}}, \bibnamefont{and}
  \bibinfo{author}{\bibfnamefont{J.~G.} \bibnamefont{Wacker}},
  \bibinfo{journal}{Phys. Rev. Lett.} \textbf{\bibinfo{volume}{113}},
  \bibinfo{pages}{171301} (\bibinfo{year}{2014}), \eprint{1402.5143}.

\bibitem[{\citenamefont{Hochberg et~al.}(2015)\citenamefont{Hochberg, Kuflik,
  Murayama, Volansky, and Wacker}}]{Hochberg:2014kqa}
\bibinfo{author}{\bibfnamefont{Y.}~\bibnamefont{Hochberg}},
  \bibinfo{author}{\bibfnamefont{E.}~\bibnamefont{Kuflik}},
  \bibinfo{author}{\bibfnamefont{H.}~\bibnamefont{Murayama}},
  \bibinfo{author}{\bibfnamefont{T.}~\bibnamefont{Volansky}}, \bibnamefont{and}
  \bibinfo{author}{\bibfnamefont{J.~G.} \bibnamefont{Wacker}},
  \bibinfo{journal}{Phys. Rev. Lett.} \textbf{\bibinfo{volume}{115}},
  \bibinfo{pages}{021301} (\bibinfo{year}{2015}), \eprint{1411.3727}.

\bibitem[{\citenamefont{Kuflik et~al.}(2016)\citenamefont{Kuflik, Perelstein,
  Lorier, and Tsai}}]{Kuflik:2015isi}
\bibinfo{author}{\bibfnamefont{E.}~\bibnamefont{Kuflik}},
  \bibinfo{author}{\bibfnamefont{M.}~\bibnamefont{Perelstein}},
  \bibinfo{author}{\bibfnamefont{N.~R.-L.} \bibnamefont{Lorier}},
  \bibnamefont{and} \bibinfo{author}{\bibfnamefont{Y.-D.} \bibnamefont{Tsai}},
  \bibinfo{journal}{Phys. Rev. Lett.} \textbf{\bibinfo{volume}{116}},
  \bibinfo{pages}{221302} (\bibinfo{year}{2016}), \eprint{1512.04545}.

\bibitem[{\citenamefont{Kuflik et~al.}(2017)\citenamefont{Kuflik, Perelstein,
  Lorier, and Tsai}}]{Kuflik:2017iqs}
\bibinfo{author}{\bibfnamefont{E.}~\bibnamefont{Kuflik}},
  \bibinfo{author}{\bibfnamefont{M.}~\bibnamefont{Perelstein}},
  \bibinfo{author}{\bibfnamefont{N.~R.-L.} \bibnamefont{Lorier}},
  \bibnamefont{and} \bibinfo{author}{\bibfnamefont{Y.-D.} \bibnamefont{Tsai}},
  \bibinfo{journal}{JHEP} \textbf{\bibinfo{volume}{08}}, \bibinfo{pages}{078}
  (\bibinfo{year}{2017}), \eprint{1706.05381}.

\bibitem[{\citenamefont{Bullock and Boylan-Kolchin}(2017)}]{Bullock:2017xww}
\bibinfo{author}{\bibfnamefont{J.~S.} \bibnamefont{Bullock}} \bibnamefont{and}
  \bibinfo{author}{\bibfnamefont{M.}~\bibnamefont{Boylan-Kolchin}},
  \bibinfo{journal}{Ann. Rev. Astron. Astrophys.}
  \textbf{\bibinfo{volume}{55}}, \bibinfo{pages}{343} (\bibinfo{year}{2017}),
  \eprint{1707.04256}.

\bibitem[{\citenamefont{Dutton et~al.}(2016)\citenamefont{Dutton, Macciò,
  Frings, Wang, Stinson, Penzo, and Kang}}]{Dutton:2015nvy}
\bibinfo{author}{\bibfnamefont{A.~A.} \bibnamefont{Dutton}},
  \bibinfo{author}{\bibfnamefont{A.~V.} \bibnamefont{Macciò}},
  \bibinfo{author}{\bibfnamefont{J.}~\bibnamefont{Frings}},
  \bibinfo{author}{\bibfnamefont{L.}~\bibnamefont{Wang}},
  \bibinfo{author}{\bibfnamefont{G.~S.} \bibnamefont{Stinson}},
  \bibinfo{author}{\bibfnamefont{C.}~\bibnamefont{Penzo}}, \bibnamefont{and}
  \bibinfo{author}{\bibfnamefont{X.}~\bibnamefont{Kang}},
  \bibinfo{journal}{Mon. Not. Roy. Astron. Soc.}
  \textbf{\bibinfo{volume}{457}}, \bibinfo{pages}{L74} (\bibinfo{year}{2016}),
  \eprint{1512.00453}.

\bibitem[{\citenamefont{Sawala et~al.}(2016)}]{Sawala:2015cdf}
\bibinfo{author}{\bibfnamefont{T.}~\bibnamefont{Sawala}} \bibnamefont{et~al.},
  \bibinfo{journal}{Mon. Not. Roy. Astron. Soc.}
  \textbf{\bibinfo{volume}{457}}, \bibinfo{pages}{1931} (\bibinfo{year}{2016}),
  \eprint{1511.01098}.

\bibitem[{\citenamefont{Wetzel et~al.}(2016)\citenamefont{Wetzel, Hopkins, Kim,
  Faucher-Giguere, Keres, and Quataert}}]{Wetzel:2016wro}
\bibinfo{author}{\bibfnamefont{A.~R.} \bibnamefont{Wetzel}},
  \bibinfo{author}{\bibfnamefont{P.~F.} \bibnamefont{Hopkins}},
  \bibinfo{author}{\bibfnamefont{J.-h.} \bibnamefont{Kim}},
  \bibinfo{author}{\bibfnamefont{C.-A.} \bibnamefont{Faucher-Giguere}},
  \bibinfo{author}{\bibfnamefont{D.}~\bibnamefont{Keres}}, \bibnamefont{and}
  \bibinfo{author}{\bibfnamefont{E.}~\bibnamefont{Quataert}},
  \bibinfo{journal}{Astrophys. J.} \textbf{\bibinfo{volume}{827}},
  \bibinfo{pages}{L23} (\bibinfo{year}{2016}), \eprint{1602.05957}.

\bibitem[{\citenamefont{Moore et~al.}(1999{\natexlab{a}})\citenamefont{Moore,
  Ghigna, Governato, Lake, Quinn, Stadel, and Tozzi}}]{Moore:1999nt}
\bibinfo{author}{\bibfnamefont{B.}~\bibnamefont{Moore}},
  \bibinfo{author}{\bibfnamefont{S.}~\bibnamefont{Ghigna}},
  \bibinfo{author}{\bibfnamefont{F.}~\bibnamefont{Governato}},
  \bibinfo{author}{\bibfnamefont{G.}~\bibnamefont{Lake}},
  \bibinfo{author}{\bibfnamefont{T.~R.} \bibnamefont{Quinn}},
  \bibinfo{author}{\bibfnamefont{J.}~\bibnamefont{Stadel}}, \bibnamefont{and}
  \bibinfo{author}{\bibfnamefont{P.}~\bibnamefont{Tozzi}},
  \bibinfo{journal}{Astrophys. J.} \textbf{\bibinfo{volume}{524}},
  \bibinfo{pages}{L19} (\bibinfo{year}{1999}{\natexlab{a}}),
  \eprint{astro-ph/9907411}.

\bibitem[{\citenamefont{Kravtsov}(2010)}]{Kravtsov:2009gi}
\bibinfo{author}{\bibfnamefont{A.~V.} \bibnamefont{Kravtsov}},
  \bibinfo{journal}{Adv. Astron.} \textbf{\bibinfo{volume}{2010}},
  \bibinfo{pages}{281913} (\bibinfo{year}{2010}), \eprint{0906.3295}.

\bibitem[{\citenamefont{Bode et~al.}(2001)\citenamefont{Bode, Ostriker, and
  Turok}}]{Bode:2000gq}
\bibinfo{author}{\bibfnamefont{P.}~\bibnamefont{Bode}},
  \bibinfo{author}{\bibfnamefont{J.~P.} \bibnamefont{Ostriker}},
  \bibnamefont{and} \bibinfo{author}{\bibfnamefont{N.}~\bibnamefont{Turok}},
  \bibinfo{journal}{Astrophys. J.} \textbf{\bibinfo{volume}{556}},
  \bibinfo{pages}{93} (\bibinfo{year}{2001}), \eprint{astro-ph/0010389}.

\bibitem[{\citenamefont{Lovell et~al.}(2012)\citenamefont{Lovell, Eke, Frenk,
  Gao, Jenkins, Theuns, Wang, White, Boyarsky, and Ruchayskiy}}]{Lovell:2011rd}
\bibinfo{author}{\bibfnamefont{M.~R.} \bibnamefont{Lovell}},
  \bibinfo{author}{\bibfnamefont{V.}~\bibnamefont{Eke}},
  \bibinfo{author}{\bibfnamefont{C.~S.} \bibnamefont{Frenk}},
  \bibinfo{author}{\bibfnamefont{L.}~\bibnamefont{Gao}},
  \bibinfo{author}{\bibfnamefont{A.}~\bibnamefont{Jenkins}},
  \bibinfo{author}{\bibfnamefont{T.}~\bibnamefont{Theuns}},
  \bibinfo{author}{\bibfnamefont{J.}~\bibnamefont{Wang}},
  \bibinfo{author}{\bibfnamefont{D.~M.} \bibnamefont{White}},
  \bibinfo{author}{\bibfnamefont{A.}~\bibnamefont{Boyarsky}}, \bibnamefont{and}
  \bibinfo{author}{\bibfnamefont{O.}~\bibnamefont{Ruchayskiy}},
  \bibinfo{journal}{Mon. Not. Roy. Astron. Soc.}
  \textbf{\bibinfo{volume}{420}}, \bibinfo{pages}{2318} (\bibinfo{year}{2012}),
  \eprint{1104.2929}.

\bibitem[{\citenamefont{Moore et~al.}(1999{\natexlab{b}})\citenamefont{Moore,
  Quinn, Governato, Stadel, and Lake}}]{Moore:1999gc}
\bibinfo{author}{\bibfnamefont{B.}~\bibnamefont{Moore}},
  \bibinfo{author}{\bibfnamefont{T.~R.} \bibnamefont{Quinn}},
  \bibinfo{author}{\bibfnamefont{F.}~\bibnamefont{Governato}},
  \bibinfo{author}{\bibfnamefont{J.}~\bibnamefont{Stadel}}, \bibnamefont{and}
  \bibinfo{author}{\bibfnamefont{G.}~\bibnamefont{Lake}},
  \bibinfo{journal}{Mon. Not. Roy. Astron. Soc.}
  \textbf{\bibinfo{volume}{310}}, \bibinfo{pages}{1147}
  (\bibinfo{year}{1999}{\natexlab{b}}), \eprint{astro-ph/9903164}.

\bibitem[{\citenamefont{de~Blok}(2010)}]{deBlok:2009sp}
\bibinfo{author}{\bibfnamefont{W.~J.~G.} \bibnamefont{de~Blok}},
  \bibinfo{journal}{Adv. Astron.} \textbf{\bibinfo{volume}{2010}},
  \bibinfo{pages}{789293} (\bibinfo{year}{2010}), \eprint{0910.3538}.

\bibitem[{\citenamefont{Spergel and Steinhardt}(2000)}]{Spergel:1999mh}
\bibinfo{author}{\bibfnamefont{D.~N.} \bibnamefont{Spergel}} \bibnamefont{and}
  \bibinfo{author}{\bibfnamefont{P.~J.} \bibnamefont{Steinhardt}},
  \bibinfo{journal}{Phys. Rev. Lett.} \textbf{\bibinfo{volume}{84}},
  \bibinfo{pages}{3760} (\bibinfo{year}{2000}), \eprint{astro-ph/9909386}.

\bibitem[{\citenamefont{Tulin and Yu}(2018)}]{Tulin:2017ara}
\bibinfo{author}{\bibfnamefont{S.}~\bibnamefont{Tulin}} \bibnamefont{and}
  \bibinfo{author}{\bibfnamefont{H.-B.} \bibnamefont{Yu}},
  \bibinfo{journal}{Phys. Rept.} \textbf{\bibinfo{volume}{730}},
  \bibinfo{pages}{1} (\bibinfo{year}{2018}), \eprint{1705.02358}.

\bibitem[{\citenamefont{Peter et~al.}(2013)\citenamefont{Peter, Rocha, Bullock,
  and Kaplinghat}}]{Peter:2012jh}
\bibinfo{author}{\bibfnamefont{A.~H.~G.} \bibnamefont{Peter}},
  \bibinfo{author}{\bibfnamefont{M.}~\bibnamefont{Rocha}},
  \bibinfo{author}{\bibfnamefont{J.~S.} \bibnamefont{Bullock}},
  \bibnamefont{and}
  \bibinfo{author}{\bibfnamefont{M.}~\bibnamefont{Kaplinghat}},
  \bibinfo{journal}{Mon. Not. Roy. Astron. Soc.}
  \textbf{\bibinfo{volume}{430}}, \bibinfo{pages}{105} (\bibinfo{year}{2013}),
  \eprint{1208.3026}.

\bibitem[{\citenamefont{Markevitch et~al.}(2004)\citenamefont{Markevitch,
  Gonzalez, Clowe, Vikhlinin, David, Forman, Jones, Murray, and
  Tucker}}]{Markevitch:2003at}
\bibinfo{author}{\bibfnamefont{M.}~\bibnamefont{Markevitch}},
  \bibinfo{author}{\bibfnamefont{A.~H.} \bibnamefont{Gonzalez}},
  \bibinfo{author}{\bibfnamefont{D.}~\bibnamefont{Clowe}},
  \bibinfo{author}{\bibfnamefont{A.}~\bibnamefont{Vikhlinin}},
  \bibinfo{author}{\bibfnamefont{L.}~\bibnamefont{David}},
  \bibinfo{author}{\bibfnamefont{W.}~\bibnamefont{Forman}},
  \bibinfo{author}{\bibfnamefont{C.}~\bibnamefont{Jones}},
  \bibinfo{author}{\bibfnamefont{S.}~\bibnamefont{Murray}}, \bibnamefont{and}
  \bibinfo{author}{\bibfnamefont{W.}~\bibnamefont{Tucker}},
  \bibinfo{journal}{Astrophys. J.} \textbf{\bibinfo{volume}{606}},
  \bibinfo{pages}{819} (\bibinfo{year}{2004}), \eprint{astro-ph/0309303}.

\bibitem[{\citenamefont{Randall et~al.}(2008)\citenamefont{Randall, Markevitch,
  Clowe, Gonzalez, and Bradac}}]{Randall:2007ph}
\bibinfo{author}{\bibfnamefont{S.~W.} \bibnamefont{Randall}},
  \bibinfo{author}{\bibfnamefont{M.}~\bibnamefont{Markevitch}},
  \bibinfo{author}{\bibfnamefont{D.}~\bibnamefont{Clowe}},
  \bibinfo{author}{\bibfnamefont{A.~H.} \bibnamefont{Gonzalez}},
  \bibnamefont{and} \bibinfo{author}{\bibfnamefont{M.}~\bibnamefont{Bradac}},
  \bibinfo{journal}{Astrophys. J.} \textbf{\bibinfo{volume}{679}},
  \bibinfo{pages}{1173} (\bibinfo{year}{2008}), \eprint{0704.0261}.

\bibitem[{\citenamefont{Kamada et~al.}(2017)\citenamefont{Kamada, Kim, and
  Sekiguchi}}]{Kamada:2017tsq}
\bibinfo{author}{\bibfnamefont{A.}~\bibnamefont{Kamada}},
  \bibinfo{author}{\bibfnamefont{H.}~\bibnamefont{Kim}}, \bibnamefont{and}
  \bibinfo{author}{\bibfnamefont{T.}~\bibnamefont{Sekiguchi}},
  \bibinfo{journal}{Phys. Rev.} \textbf{\bibinfo{volume}{D96}},
  \bibinfo{pages}{016007} (\bibinfo{year}{2017}), \eprint{1704.04505}.

\bibitem[{\citenamefont{Kamada et~al.}(in prep)\citenamefont{Kamada, Kim, Kim,
  and Sekiguchi}}]{KKKSip}
\bibinfo{author}{\bibfnamefont{A.}~\bibnamefont{Kamada}},
  \bibinfo{author}{\bibfnamefont{H.~J.} \bibnamefont{Kim}},
  \bibinfo{author}{\bibfnamefont{H.}~\bibnamefont{Kim}}, \bibnamefont{and}
  \bibinfo{author}{\bibfnamefont{T.}~\bibnamefont{Sekiguchi}}
  (\bibinfo{year}{in prep}).

\bibitem[{\citenamefont{Kamada et~al.}(2013)\citenamefont{Kamada, Yoshida,
  Kohri, and Takahashi}}]{Kamada:2013sh}
\bibinfo{author}{\bibfnamefont{A.}~\bibnamefont{Kamada}},
  \bibinfo{author}{\bibfnamefont{N.}~\bibnamefont{Yoshida}},
  \bibinfo{author}{\bibfnamefont{K.}~\bibnamefont{Kohri}}, \bibnamefont{and}
  \bibinfo{author}{\bibfnamefont{T.}~\bibnamefont{Takahashi}},
  \bibinfo{journal}{JCAP} \textbf{\bibinfo{volume}{1303}}, \bibinfo{pages}{008}
  (\bibinfo{year}{2013}), \eprint{1301.2744}.

\bibitem[{\citenamefont{Ir$\check{\rm s}$i$\check{\rm c}$
  et~al.}(2017)}]{Irsic:2017ixq}
\bibinfo{author}{\bibfnamefont{V.}~\bibnamefont{Ir$\check{\rm s}$i$\check{\rm
  c}$}} \bibnamefont{et~al.}, \bibinfo{journal}{Phys. Rev.}
  \textbf{\bibinfo{volume}{D96}}, \bibinfo{pages}{023522}
  (\bibinfo{year}{2017}), \eprint{1702.01764}.

\bibitem[{\citenamefont{Yèche et~al.}(2017)\citenamefont{Yèche,
  Palanque-Delabrouille, Baur, and du~Mas~des Bourboux}}]{Yeche:2017upn}
\bibinfo{author}{\bibfnamefont{C.}~\bibnamefont{Yèche}},
  \bibinfo{author}{\bibfnamefont{N.}~\bibnamefont{Palanque-Delabrouille}},
  \bibinfo{author}{\bibfnamefont{J.}~\bibnamefont{Baur}}, \bibnamefont{and}
  \bibinfo{author}{\bibfnamefont{H.}~\bibnamefont{du~Mas~des Bourboux}},
  \bibinfo{journal}{JCAP} \textbf{\bibinfo{volume}{1706}}, \bibinfo{pages}{047}
  (\bibinfo{year}{2017}), \eprint{1702.03314}.

\bibitem[{\citenamefont{Valli and Yu}(2017)}]{Valli:2017ktb}
\bibinfo{author}{\bibfnamefont{M.}~\bibnamefont{Valli}} \bibnamefont{and}
  \bibinfo{author}{\bibfnamefont{H.-B.} \bibnamefont{Yu}}
  (\bibinfo{year}{2017}), \eprint{1711.03502}.

\bibitem[{\citenamefont{Sekiguchi and Tashiro}(2014)}]{Sekiguchi:2014wfa}
\bibinfo{author}{\bibfnamefont{T.}~\bibnamefont{Sekiguchi}} \bibnamefont{and}
  \bibinfo{author}{\bibfnamefont{H.}~\bibnamefont{Tashiro}},
  \bibinfo{journal}{JCAP} \textbf{\bibinfo{volume}{1408}}, \bibinfo{pages}{007}
  (\bibinfo{year}{2014}), \eprint{1401.5563}.

\bibitem[{\citenamefont{Gilman et~al.}(2017)\citenamefont{Gilman, Birrer, Treu,
  and Keeton}}]{Gilman:2017voy}
\bibinfo{author}{\bibfnamefont{D.}~\bibnamefont{Gilman}},
  \bibinfo{author}{\bibfnamefont{S.}~\bibnamefont{Birrer}},
  \bibinfo{author}{\bibfnamefont{T.}~\bibnamefont{Treu}}, \bibnamefont{and}
  \bibinfo{author}{\bibfnamefont{C.~R.} \bibnamefont{Keeton}}
  (\bibinfo{year}{2017}), \eprint{1712.04945}.

\bibitem[{\citenamefont{Ade et~al.}(2016)}]{Ade:2015xua}
\bibinfo{author}{\bibfnamefont{P.~A.~R.} \bibnamefont{Ade}}
  \bibnamefont{et~al.} (\bibinfo{collaboration}{Planck}),
  \bibinfo{journal}{Astron. Astrophys.} \textbf{\bibinfo{volume}{594}},
  \bibinfo{pages}{A13} (\bibinfo{year}{2016}), \eprint{1502.01589}.

\end{thebibliography}

\onecolumngrid
\appendix
\newpage

\end{document}